\documentclass[preprint,pre,nofootinbib,superscriptaddress,aps]{revtex4-1}
\usepackage{amssymb}
\usepackage{hyperref}
\usepackage{amsmath}
\usepackage{graphicx}
\usepackage{color}

\begin{document}
\title{Accelerated Stochastic Sampling of Discrete Statistical Systems}
\author{Zsolt \surname{Bertalan}}
\email{zsolt@stat.phys.titech.ac.jp}
\author{Hidetoshi \surname{Nishimori}}
\affiliation{Department of Physics, Tokyo Institute of Technology, Oh-okayama, Meguro-ku, Tokyo 152-8551, Japan
}
\author{Henri \surname{Orland}}
\affiliation{Institut de Physique Th\'eorique
CEA, IPhT and CNRS, URA 2306
 F-91191 Gif-sur-Yvette, France
}
\date{\today}
\begin{abstract}
We propose a method to reduce the relaxation time towards equilibrium in stochastic sampling of complex energy landscapes in statistical systems with discrete degrees of freedom by generalizing the platform previously developed for continuous systems. The method starts from a master equation, in contrast to the Fokker-Planck equation for the continuous case. The master equation is transformed into an imaginary-time Schr\"odinger equation. The Hamiltonian of the Schr\"odinger equation is modified by adding  a projector to its known ground state. We show how this transformation decreases the relaxation time and propose a way to use it to accelerate simulated annealing for optimization problems. We implement our method in a simplified kinetic Monte Carlo scheme and show an acceleration by an order of magnitude in simulated annealing of the symmetric traveling salesman problem. Comparisons of simulated annealing are made with the exchange Monte Carlo algorithm for the three-dimensional Ising spin glass. Our implementation can be seen as a step toward accelerating the stochastic sampling of generic systems with complex landscapes and long equilibration times.    
\end{abstract}
\maketitle

\section {Introduction}\label{sec:intro}
Numerical sampling of rugged energy landscapes is notoriously difficult \citep{hartmann-rieger}. Transition rates between two states are exponential functions of 
%their energy difference
the energy barrier between them
divided by the temperature. The sampling of complex systems is a daunting task because there are many states of comparable energies separated by large barriers. One of the most widespread sampling methods at finite temperatures is the Monte Carlo method, where one follows one or more trajectories of a virtual Brownian particle as it moves through the state space. To sample a system at a given temperature, there exists a plethora of approaches, and  amongst many others particularly notable are the standard (Metropolis) \citep{metropolis} and the kinetic Monte Carlo \citep{voter} methods. At low temperatures these stochastic schemes tend to take a long time before a satisfactory result is reached. If one is interested in the behavior of a specific problem in the low-temperature limit, the common method is simulated annealing \citep{KGV}. If, however, one samples systems where below a certain temperature the state space splits into regions which are separated by huge barriers, like for example in spin glass systems, simulated annealing does not always lead to satisfactory results. Examples of accepted solutions to this problem include the exchange Monte Carlo method \citep{XMC} and the population annealing \citep{popa}.
Although good results are obtainable for most systems using these techniques, relaxation time still stays a crucial factor. For the protein folding problem \citep{pfolding} for example, the longest simulations available on present day computers are still far from the equilibrium distributions.  A possible step in resolving the problem of over-long relaxation times was proposed in Ref. \citep{orland} where a method to accelerate the sampling of continuous systems was introduced. The basic idea is to rewrite the Fokker-Planck equation, which describes the time evolution of the probability distribution for continuous systems, into an imaginary-time Schr\"odinger equation, for which one artificially introduces an energy gap between the ground state and the first excited state. Then the relaxation time, which is proportional to the inverse of the energy gap, is reduced.

In the present paper, we extend the idea of Ref. \cite{orland} in Sec. \ref{sec:method} to discrete systems and implement it into a stochastic sampling scheme. Next, in Sec. \ref{sec:results}, we analyze how our method can be used to improve the performance of simulated annealing of the traveling salesman problem \cite{reinelt} and the three-dimensional Ising spin glass. In the case of the traveling salesman problem we find that the simulated annealing is significantly accelerated and the modified sampling finds the approximately optimal solution much faster than unmodified simulations. Our method also leads to improvements for the three-dimensional Gaussian Ising spin glass. Sec. \ref{sec:conclusio} concludes this paper.

\section{Accelerated Sampling}\label{sec:method}
In this section we extend idea of Ref. \citep{orland} to discrete systems and discuss its implementation into the kinetic Monte Carlo algorithm. We furthermore introduce a simplified version of the kinetic Monte Carlo scheme to speed up the calculations and save computational resources.
\subsection{Widening the Gap}
The master equation whose transition rates $w_{ab}$ fulfill the detailed balance condition reads,
\begin{equation}\label{eqn:masta}
\frac{d P_a}{dt}=\sum_b{w_{ab}P_b}-\sum_b{w_{ba}P_a}\quad\quad\left(\frac{w_{ba}}{w_{ab}}=e^{-\beta(E_b-E_a)}\right).
\end{equation}
If we set $P_a=f_aQ_a$, with $f_a=e^{-\beta E_a/2}/\sqrt{Z}$, where $Z=\sum_b \exp(-\beta E_b)$ is the partition function of the system and $E_a$ is the energy of state $a$, we get 
\begin{equation}
\frac{dQ_a}{dt}=\sum_b{\frac{f_b}{f_a}w_{ab}Q_b}-\sum_bw_{ba}Q_a\equiv -\sum_b H_{ab}Q_b\label{eqn:HAMab}.
\end{equation}
Since $H_{ab}$, as defined in Eq. (\ref{eqn:HAMab}), is a real, symmetric matrix, we call it a Hamiltonian and thus, Eq. (\ref{eqn:HAMab}) may be regarded as an imaginary-time Schr\"odinger equation. It has a zero eigenvalue with eigenvector $f_b$, $\sum_bH_{ab}f_b=0$. This follows directly from the definitions of $f_b$ and $H_{ab}$ and the detailed balance condition in Eq. (\ref{eqn:masta}). The lowest eigenvalue of $H_{ab}$ is therefore zero as guaranteed by the Perron-Frobenius theorem. 

Following the idea in Ref. \cite{orland}, we make the transformation $H_{ab}\rightarrow H_{ab}+\lambda(\delta_{ab}-P^0_{ab})$, where $P^0_{ab}$ is the projector to the state of zero eigenvalue, which is expected to shorten the relaxation time toward equilibrium. The spectrum of $H_{ab}$ is then shifted by $\lambda$ for all states except the ground state. It is easy to see that the matrix elements of the projector are $P^0_{ab}={f_af_b}$, since $\sum_bP^0_{ab}f_b=f_a\sum_b f^2_b=f_a\sum_b \exp(-\beta E_b)/Z=f_a$ and the eigenvectors of $H_{ab}$ corresponding to different eigenvalues are orthogonal. Therefore, the imaginary-time Schr\"odinger equation with the modified Hamiltonian is
\begin{eqnarray}\label{eqn:Qlambda}
\frac{dQ_a}{dt}=\sum_b{\left(\frac{f_b}{f_a}w_{ab}+\lambda \frac{f_af_b}{Z} \right)Q_b}-Q_a\left(\lambda+\sum_bw_{ba}\right).
\end{eqnarray}
The solutions of this equation are,
\begin{equation}
Q_a(t)=Q_a^{(0)}+Q_a^{(1)}e^{-t (\epsilon_1+\lambda)}+Q_a^{(2)}e^{-t (\epsilon_2+\lambda)}+\cdots\label{eqn:Qsol}
\end{equation}
where the $\{Q^{(n)}\}$ and $\{\epsilon_n\}$ are the eigenvectors and corresponding eigenvalues of $H_{ab}$, respectively.

Making use of the relation $P_a=f_aQ_a$, we can translate Eq. (\ref{eqn:Qlambda}) into the modified master equation 
\begin{equation}
\dfrac{dP_a}{dt}=\sum_b \left(w_{ab}+\lambda \dfrac{e^{-\beta E_a}}{Z}\right)P_b
-P_a \left(\lambda+\sum_b w_{ba}\right)\end{equation} 
and deduce its solution from Eq. (\ref{eqn:Qsol}) as 
\begin{equation}\label{eqn:Pt}
P_a(t)=\frac{e^{-\beta E_a} }{Z}+P_a^{(1)}e^{-t(\epsilon_1+\lambda)}+P_a^{(2)}e^{-t(\epsilon_2+\lambda)}+\cdots.
\end{equation}
For large times this decays to
\begin{equation}
P_a=\frac{e^{-\beta E_a}}{Z},
\end{equation}
i.e. the Boltzmann weight, as expected. However, as can be seen from Eq. (\ref{eqn:Pt}), the relaxation is faster than the case where $\lambda$ is absent.

\subsection{Implementation}\label{sec:implementation}
In Ref. \cite{orland}, a similar idea was tested for continuous systems using a diffusion Monte Carlo calculation. In the present discrete case, a straightforward implementation of our method is through the kinetic Monte Carlo algorithm \citep{voter}. Let us first describe this Monte Carlo method on the original master equation 
(\ref{eqn:masta}) in order to make is clear what parts need modifications to accommodate the $\lambda$-term in Eq. (\ref{eqn:Qlambda}). The idea is to try to generate time `trajectories' of the system among its various available states in such a way as to satisfy the master equation.
Assume the system is in a given state $a$. The rate (or probability per unit time) at which the system will escape from $a$ to any available state $b$ is given by the second term on the right-hand side of Eq. (\ref{eqn:masta}). In other words, it is equal to $\Gamma_a =\sum_b w_{ba}$. Therefore, the probability distribution of the escape-time from $a$ is given by the Poisson distribution $P^{\textrm{esc}}_a (t) = \Gamma_a e^{-\Gamma_a t}$. The time at which the system will leave state $a$ can thus be drawn from this distribution. The probability for the system to go from $a$ to a state $b$ is obviously given by the ratio ${w_{ba}}/{\Gamma_a}$. 
 
 The practical implementation of the algorithm goes as follows. Assume that the system is in state $a$ at time $t_n$:
 \begin{itemize} 
 \item The system will make a random transition out of state $a$ at a time  $t_{n+1} = t_n + \tau$ where $\tau$ is drawn from the distribution  $P^{\textrm{esc}}_a (\tau) = \Gamma_a e^{-\Gamma_a \tau}$. In practice, one draws a number $r$ uniformly distributed between 0 and 1 and takes $\tau = -\frac{1}{\Gamma_a} \log (-r)$.
 \item The state $b$ to which the system will make the transition is chosen with probability ${w_{ba}}/{\Gamma_a}$. A simple way to do this is to draw all individual probabilities consecutively until eventually they fill up the interval $[0,1]$.  Then one draws a uniform random number $r_1$ between 0 and 1, and the state $b$ to which the system jumps is the one indexed by the transition probability which corresponds to $r_1$ on the interval [0,1].
 \end{itemize}
 
By generating many such trajectories, one generates probability distributions $P_a(t)$ which stochastically satisfy the master equation.
Let us note that all trajectories generated this way are \textit{statistically independent}, and can thus be used to compute averages.

We now turn to an implementation of the kinetic Monte Carlo when one introduces the parameter $\lambda$. Let us first expand unity as 
\begin{equation}\label{eqn:unity}
1=\sum_{c(a)} \frac{e^{-\beta E_c}}{Z_a}\quad\quad \left(Z_a=\sum_{c(a)} e^{-\beta E_c}\right),
\end{equation}
where the sum is over the states $\{c(a)\}$ accessible from a given state $a$. If all states are accessible in principle, which is the case if the system has no intrinsic dynamics, like the Ising model, then we restrict $\{c(a)\}$ to some subset depending on $a$, for example the nearest neighbors. Next, insert Eq. (\ref{eqn:unity}) into the outgoing part of the master equation, to obtain
\begin{equation}\label{eqn:out}
\lambda +\sum_{c(a)} w_{ca}=\sum_{c(a)} \left(w_{ca}+\lambda\frac{e^{-\beta E_c}}{Z_a}\right)\equiv\sum_{c(a)}w^{\lambda}_{ca}.
\end{equation}
Note that, even if the physical considerations do not allow non-local moves, we can define transition probabilities $w^{\lambda}_{ca}=\lambda e^{-\beta E_c}/Z_a$ for such moves and thus incorporate non-local states into the list $\{c(a)\}$ in a very straightforward manner.

The introduction of the term $Z_a$ in Eq. (\ref{eqn:unity}) and Eq. (\ref{eqn:out}) makes the present implementation unsuitable to reproduce finite-temperature properties. To show this fact, let us first recall that a frequently-used transition probability is the heat bath (or Glauber) method
\begin{equation}\label{glauber_trprob}
w_{ca}=\frac{e^{-\beta E_c}}{e^{-\beta E_c }+e^{-\beta E_a}},
\end{equation}   
which trivially fulfills the detailed balance condition. However, the modified master equation is not symmetric in its outgoing and incoming parts. The kinetic Monte Carlo algorithm uses only the outgoing part for sampling, and therefore, adding the $\lambda$ terms tilts the detailed balance in favor of the energetically lower lying states. To see this, we write the detailed balance condition as
\begin{equation}
\frac{w^{\lambda}_{ca}}{w^{\lambda}_{ac}}=\frac{\displaystyle{w_{ca}+\lambda\frac{\exp(-\beta E_c)}{Z_a}}} {\displaystyle{w_{ac}+\lambda\frac{\exp(-\beta E_a)}{Z_c}}}.
\end{equation}
Inserting the transition rate (\ref{glauber_trprob}) into the above equation, we get
\begin{equation}\label{eqn:tilted_db}
\frac{w^{\lambda}_{ca}}{w^{\lambda}_{ac}}=
\frac
{e^{-\beta E_c}(1+\lambda [e^{-\beta E_a}+e^{-\beta E_c }] / Z_a)}
{e^{-\beta E_a}(1+\lambda [e^{-\beta E_a}+e^{-\beta E_c }] /Z_c)}=e^{-\beta(E_c-E_a)}\frac{Z_c}{Z_a}\frac{Z_a+\lambda [e^{-\beta E_a}+e^{-\beta E_c }]}{Z_c+\lambda [e^{-\beta E_a}+e^{-\beta E_c }]}.
\end{equation}
We see clearly that the detailed balance in its conventional form is not satisfied. If we recall that $Z_k=\sum_{m(k)}e^{-\beta E_m}$, which represents the sum of Boltzmann factors of the subset $\{m(k)\}$ of states accessible from state $k$, then we can conclude that if, $Z_c<Z_a$, there are more energetically lower lying states available from $a$ than from $c$. This would suggest that in this implementation (Eq. (\ref{eqn:out})) the addition of $\lambda$ indeed tilts the detailed balance in favor of states from where more lower lying states are accessible. This changes the finite temperature values of physical observables when compared to calculations made with $\lambda=0$ at the same temperature. Nevertheless, when we are interested in the ground state solution, we do not have to worry about such finite temperature differences.    

There may be other implementations that do not use state- or temperature-dependent renormalizations of $\lambda$ and thus allow us to keep detailed balance. However, we reserve the finite temperature case for future studies. 

Using the language and example of the traveling salesman problem (see Sec. \ref{sec:models}) we now introduce a simplified version of the kinetic Monte Carlo scheme. Sampling all nearest neighbors of a given tour, as needed for the kinetic Monte Carlo, is neither efficient nor feasible. At every Monte Carlo step we would have to calculate $N(N-1)/2$ transition rates, where $N$ is the number of cities in the map. For example, if $N\approx1000$ we would have to make about $5\cdot 10^5$ calculations. Therefore, we choose just a certain number of nearest neighbors and generate another number of non-local states to build our list $\{c(a)\}$. This is possible since the traveling salesman problem has no intrinsic dynamics, and thus, all other states are accessible from any given state. By taking a set of nearest neighbor states and a set of non-local states as possible jumps available, we sample in effect the local structure as well as `far' away states and can in this fashion overcome large barriers. We can either choose to sample less local or non-local states than in the case of the kinetic Monte Carlo. In such a way the computational cost is reduced.
\begin{figure}[ht]
\includegraphics[width=\textwidth]{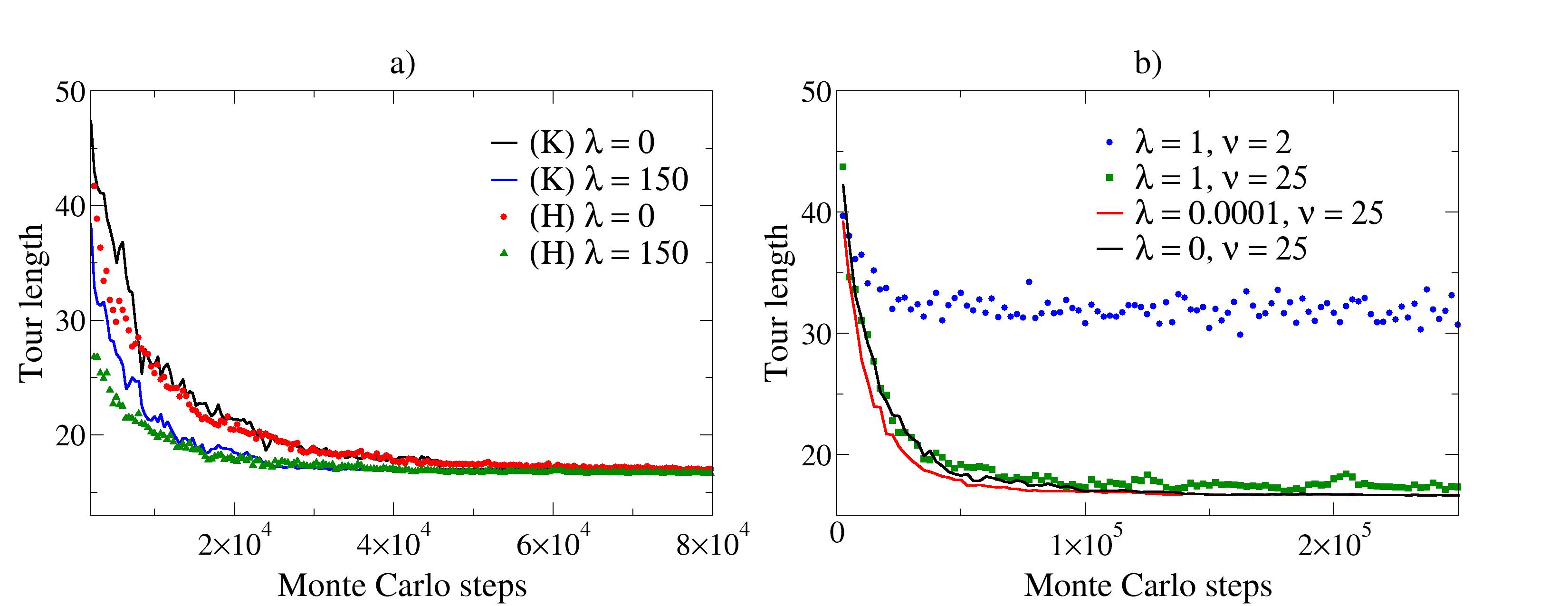}
\caption{(Color online) a) Simulated annealing of the traveling salesman map `gr229' with 229 cities. Comparison of the full kinetic Monte Carlo (K) and the simplified (H) method. b) Investigation of the simplified method with $\nu$ non-local and no local trials. See the text for details.}\label{KMCHY} 
\end{figure}

Before performing full-scale computations, we run preliminary simulations to check if the simplified method actually works and investigate what values of various parameters are to be used in practical calculations. In Fig. \ref{KMCHY} a) we compare the full kinetic Monte Carlo (labeled K) and the simplified method (labeled H). We perform simulated annealing calculations of a small traveling salesman map, `gr229' \cite{reinelt}, with 229 cities. The annealing schedule is chosen to be step-wise growing (see Fig. \ref{fig:schedule} for comparison), starting from an already low $\beta_0=40$, the inverse temperature is increased by $\delta\beta$ every $2000$ Monte Carlo steps. For the simplified method, we chose 230 local states and 230 non-local states per Monte Carlo step at random to build the list $\{c(a)\}$, and sample all nearest neighbors plus 230 non-local states for the full kinetic Monte Carlo. The simplified method outperforms the full kinetic Monte Carlo, as the simplified method finds a shorter tour within the investigated time window. 

In Fig. \ref{KMCHY} b) we investigate the parameter dependence of the simplified method. We look at the case where we allow no nearest neighbor hops and sample $\nu$ non-local states per Monte Carlo step to see the effect of adding $\lambda$ to the transition rates while taking too few states to build the list $\{c(a)\}$. We see that for $\nu=25$ and very small $\lambda=0.0001$ (red line) there is a slight visible deviation to shorter tours from the simulation with $\lambda=0$ (black line), while a larger value of $\lambda=1$ (green squares) affects the sampling in a negative way. As an extreme case we take only $\nu=2$ non-local states and $\lambda=1$ (blue circles), and see that the system does not relax. The reason for this failure at $\lambda=1$, $\nu=2$ is that, with adding $\lambda$ to the transition rates, jumps are facilitated. Since we are looking at only two possible jumps, at some point both transitions will become approximately equally likely and the system fails to relax. We see that, if too few states to which jumps are possible are chosen, the sampling is influenced in an undesirable way.

  We use this simplified version of the kinetic Monte Carlo algorithm also for the Ising model, since for larger system sizes calculating the Hamiltonian and exponentiating it $L^3$ times (there are $L^3$ nearest neighbors to a given state) becomes quickly very time-consuming. As a rule of thumb, we sample $O(L)$, instead of $L^3$, local states per Monte Carlo step.

\section{Results}\label{sec:results}
In this section we begin with a short review of the models we used for our calculations to establish the terminology. Then we show the results of our simulations for different scenarios.

\subsection{The Models}\label{sec:models}
\begin{figure}[ht]
	\centering
		\includegraphics[width=0.90\textwidth]{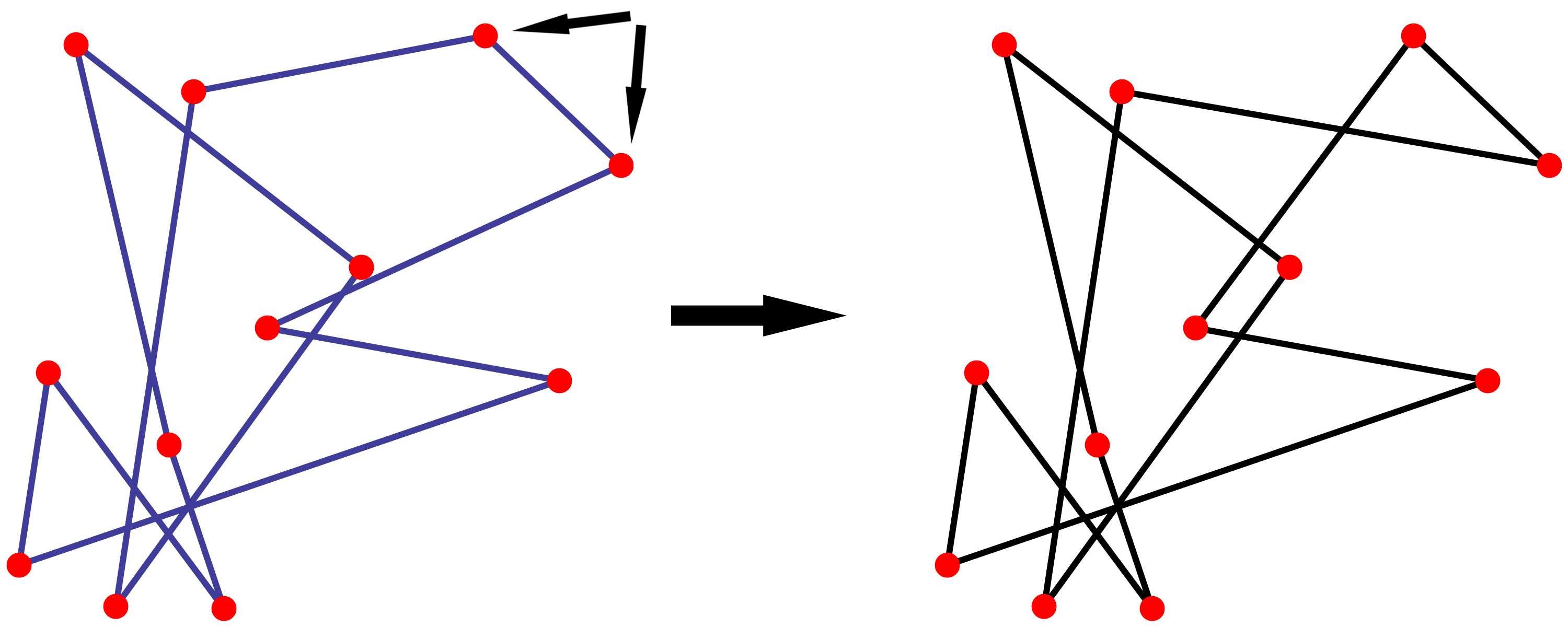}
\caption{(Color online) Illustration of a nearest neighbor hop of the traveling salesman problem. The right tour differs from the left one by the exchange of the points indicated. }
	\label{fig:TSP}
\end{figure}

We applied our method to two models. First we treat instances of the symmetric traveling salesman problem. Given a set of coordinates $M=\{(x_i,y_i)\}$ on a two-dimensional plane, the task is to find the shortest closed path, called a tour, connecting all points while traversing each point only once. Therefore, tours are ordered lists of the coordinates $M$, giving the rule in which order to visit the coordinates. We take the distance between two points $i$ and $j$ on a tour to be Euclidean $d_E(i,j)=\sqrt{(x_j-x_i)^2+(y_j-y_i)^2}$ so that the total length of the tour is calculated as the sum of all segments $E(r)=\sum d_E(i,j)$. To formulate this as a pseudo-physical problem, we identify the tours as the states, the tour length as the Hamiltonian and we choose nearest neighbor hoppings as the local dynamics. Nearest neighbors of a tour $r$ are defined as tours $s$ differing by the exchange of two points, i.e. having Hamming distance $d_H$ of two to $r$, $N(r)=\{s|d_H(r,s)=2\}$, see Fig. \ref{fig:TSP}. 

With only nearest neighbor hoppings it is hardly possible to solve the traveling salesman problem to optimality
\citep{johnson}. We can, however, efficiently choose non-local hoppings \citep{lin}, reversals or transport of tour segments.

The second model we treat is the three-dimensional Ising model on the cubic lattice of linear size $L$, with random bonds. The energy of a given spin configuration $S^a$ is calculated by $ E_a=-\sum  J_{ij}S_i^aS^a_j-h\sum S_i^a$, where the spins take values $S_i^a \in \{-1,1\}$. The bonds $J_{ij}$ are quenched random numbers drawn from the distribution $P(J_{ij})=\exp(-{J_{ij}^2}/{2J^2})/{\sqrt{2\pi J^2}}$, which we will call the Gaussian Ising spin glass, and $h$ is an external field.
We take only nearest neighbor jumps for the dynamics and use no non-local cluster flips \cite{wolff}, because we want to compare our method to the exchange Monte Carlo which gives satisfactory results with local sampling only. The nearest neighbors of a spin configuration $S^a$ are the configurations $S^b$ which differ from $S^a$ by a single spin flip, $d_H(S_a,S_b)=1$.

\subsection{Simulated Annealing}
Simulated annealing \citep{KGV} relies on one hand on a stochastic sampling scheme and on the other on an annealing schedule $\beta(n)$, the rule of increase of the inverse temperature with iterations. We first show on the  wiggly harmonic potential discussed in Ref. \citep{shinokaba} what the effect of introducing $\lambda$ is and then discuss the application of our method to simulated annealing. In this spirit we then apply simulated annealing to the traveling salesman problem and the three-dimensional Gaussian Ising spin glass. 
\subsubsection{Wiggly Harmonic Potential}
\begin{figure}[ht]
\centering
\includegraphics[width=0.5\textwidth]{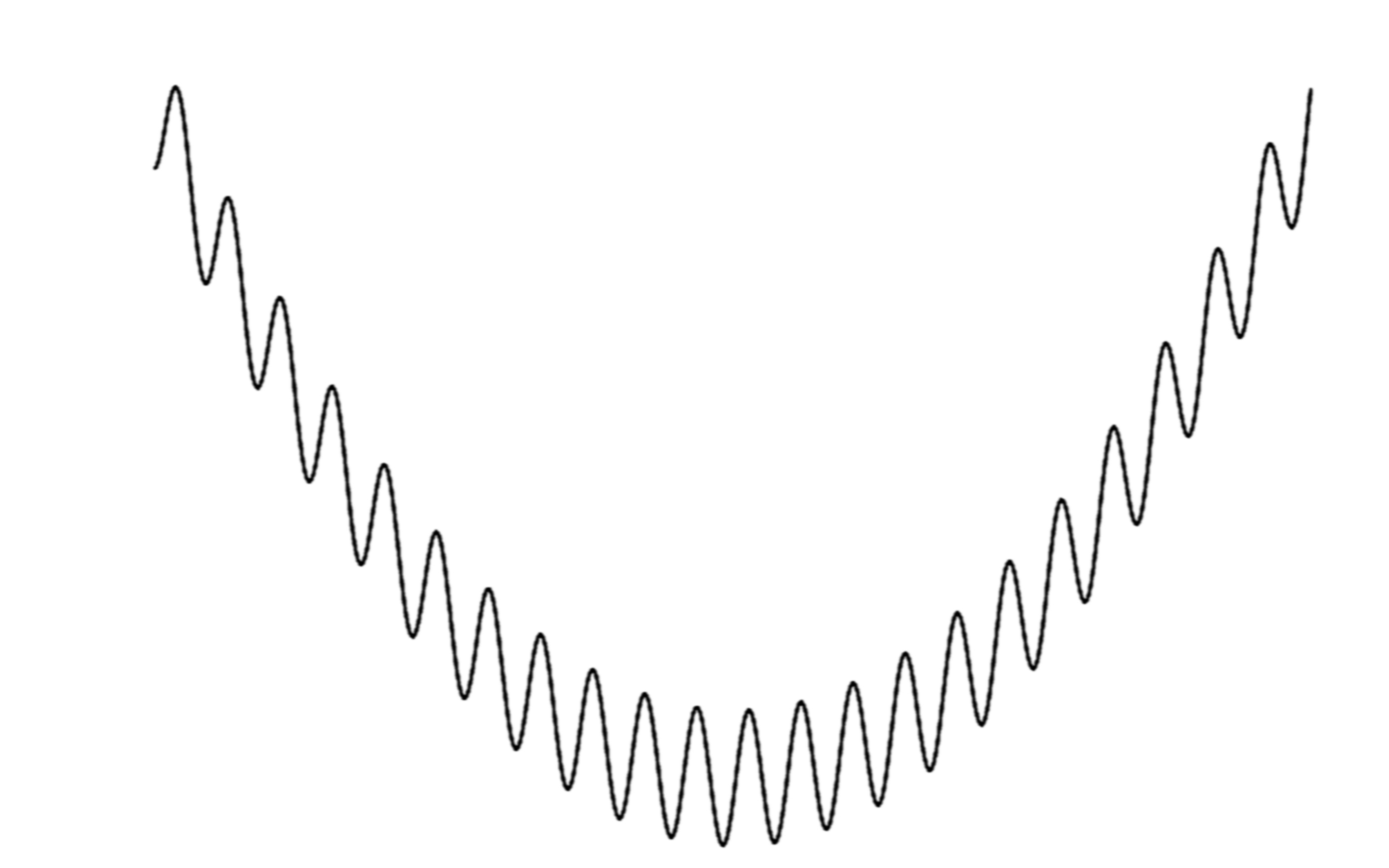} 
\caption{The landscape of the wiggly harmonic potential.}\label{fig:wiggly}
\end{figure}

Consider a one-dimensional system where a number of local minima is evenly distributed over the large basin $U=x^2/2$, as in Fig. \ref{fig:wiggly}. The distance $a$ between two neighboring minima is kept constant. The random walk of a point in this scenario is governed by the master equation describing the hopping between neighboring sites $k-1$, $k$ and $k+1$:
\begin{eqnarray}
\frac{d P_k}{dt}=e^{-B\beta}P_{k+1} +e^{-(B+\Delta_{k-1})\beta}P_{k-1}-(e^{-(B+\Delta_{k})}+e^{-B\beta}) P_{k},
\end{eqnarray}
where $B$ is the height of the barrier separating two minima and $\Delta_k= {a^2((k+1)^2-k^2)}/{2}$. When we take the continuum limit $a\ll 1$, the coarse-grained master equation becomes the Fokker-Planck equation \cite{shinokaba}
\begin{eqnarray}
\frac{\partial P(x,t)}{\partial t}&=&D\frac{\partial^2 P(x,t)}{\partial x^2}+\beta D\frac{\partial [x P(x,t)]}{\partial x}\label{fokkerplanck}
\end{eqnarray}
with diffusion constant $D=e^{-B\beta}/a^2$. The fastest annealing schedule $\beta(t)$ that minimizes the average energy
\begin{equation}
y(t)=\int dx U(x) P(x,t)
\end{equation}
 is given by $\beta(t)=\ln t/B$ \citep{shinokaba}, which coincides with the generic bound for convergence to reach the global minimum \citep{geman}. With this schedule the average energy decays as $y\sim 1/\ln t$. `Faster' schedules than this $\ln t/B$ do not further minimize the average energy.

\begin{figure}[ht]
\centering
\includegraphics[width=0.35\textwidth]{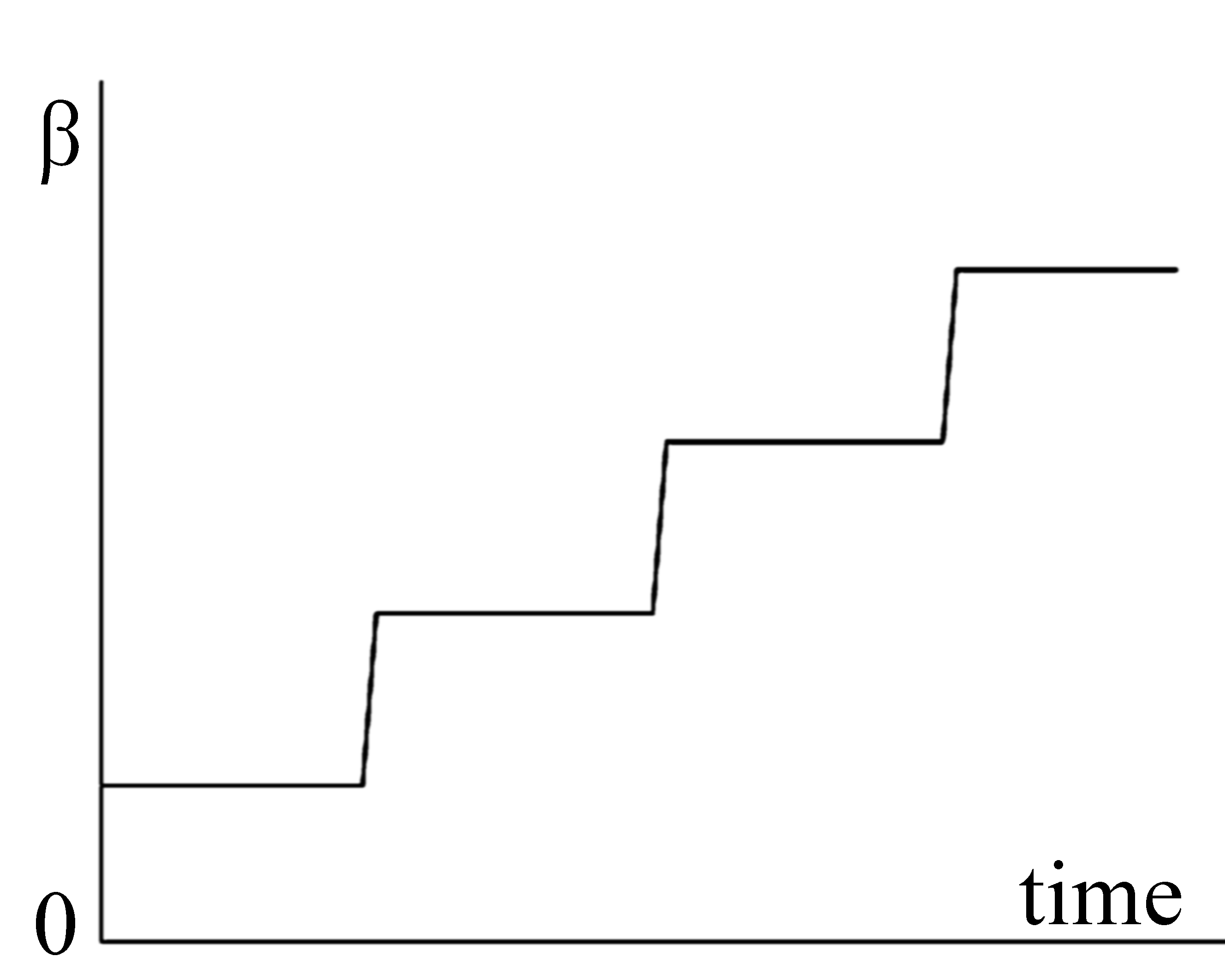} 
\caption{Step-wise growing schedule. The inverse temperature is increased by a constant amount in regular intervals.}\label{fig:schedule}
\end{figure}

 We identify the cause for this inverse-log law for the wiggly harmonic potential as an instability in the associated imaginary-time Schr\"odinger equation. Then we will propose a way how a step-wise growing schedule, see Fig. \ref{fig:schedule}, together with the considerations of Sec. \ref{sec:method} can improve the performance. The average energy in this improved case scales as $y(t_i)=\langle E_i\rangle \sim 1/\delta i$, where $i$ is the time step index and $\delta$ is some constant, which implies $y(t)\sim 1/t$. 

To rewrite Eq. (\ref{fokkerplanck}) as an imaginary-time Schr\"odinger equation, let us set $P(x,t)=e^{-\beta(t) U(x)/2}\psi(x,t)$ to get
\begin{eqnarray}
-\frac{\partial \psi(x,t)}{\partial t}&=&-D\frac{\partial^2 \psi(x,t)}{\partial x^2}+\left[D\left(\frac{\beta}{2}U'(x)\right)^2-D\frac{\beta}{2}U''(x)-\frac{\dot\beta U(x)}{2}\right]\psi(x,t)\nonumber\\
&=&-D\frac{\partial^2 \psi(x,t)}{\partial x^2}+\left[D\left(\frac{\beta}{2}x\right)^2-D\frac{\beta}{2}-\frac{\dot\beta x^2}{4}\right]\psi(x,t)\label{psi},
\end{eqnarray}
where $\dot\beta=d\beta/dt$ is the time-derivative of the schedule.
We can draw on an analogy with the quantum harmonic oscillator ($\hbar=1$) to have $m=1/2D$, $\omega=D^{1/2}\left(D\beta^2-\dot\beta\right)^{1/2}$.
For a meaningful analysis we require that the separation of energy levels, which is proportional to the frequency $\omega$, be greater than or at least equal to zero,
\begin{eqnarray}
D\beta^2-\dot\beta&\geq&0
\end{eqnarray}
or
\begin{eqnarray}
e^{-B\beta}\beta^2&\geq&a^2\dot\beta\label{stability}.
\end{eqnarray}
This relation is asymptotically only fulfilled for functions which grow like $\ln t$ or slower than that. To see this we first integrate Eq. (\ref{stability}) and have
\begin{equation}
t\geq\int \frac{d\beta}{\beta^2}a^2e^{B\beta}.
\end{equation}
We take the large $\beta$ limit and ignore $(a/\beta)^{2}$ compared to $e^{B\beta}$ to get the usual $\ln t/B\agt\beta$ schedule restriction.  For $e^{-B\beta}(\beta/a)^2=\dot\beta$ the energy levels coalesce and Eq. (\ref{psi}) becomes a free diffusion equation with time dependent, monotonically decreasing diffusion coefficient. Schedules which do not fulfill the relation (\ref{stability}) have an imaginary $\omega$ and do not lead to a decaying solution.

 To circumvent this problem we employ the method proposed in \citep{orland} (see also Sec. \ref{sec:method}) for the accelerated sampling of Boltzmann distributions. If we choose the schedule like in Fig. \ref{fig:schedule}, $\beta(t)=\sum\beta_i\Theta(t-t_i)$, where $\Theta(t)$ is the Heaviside step function, then for intermediate $t_i<t<t_{i+1}$ times, the Schr\"odinger equation (\ref{psi}) reads 
\begin{eqnarray}
-\frac{\partial\psi(x,t)}{\partial t}=-D_i\frac{\partial^2\psi(x,t)}{\partial x^2}+D_i\left[\left(\frac{\beta_i}{2}x\right)^2-\frac{\beta_i}{2}\right]\psi(x,t)= H_i\psi\label{psiho},
\end{eqnarray}
where $\beta_i=\sum_{j=0}^i\beta_j$ and $D_i=D(\beta_i)=e^{-B\beta_i}/a^2$.
Again, in correspondence with a harmonic oscillator, we obtain
\begin{eqnarray}
m_i&=&\frac{1}{2D_i}\\
\omega_i&=&\beta_iD_i\\
E_i^{(n)}&=&n\omega_i=n\beta_iD_i=n\beta_i a^{-2} e^{-B\beta_i}\label{eqn:E_i}
\end{eqnarray}
and denote the eigenfunction of $H_i$ in Eq. (\ref{psiho}) to the eigenvalue $E_i^{(n)}$ as $\phi_i^{(n)}$. 
We now employ the sudden approximation \citep{schwabl}, which uses the fact that, if the system changes too quickly, the wave function cannot follow and we can use the new Hamiltonian in the time evolution operator with the previous wave function as the initial condition. Let us assume that a particle is in the ground state of the Hamiltonian at time step $i-1$, $\phi_{i-1}^0$. Then just after the jump to the next time step $i$, the wave function of the particle can be described by
\begin{eqnarray}
\psi(t)=e^{-H_i (t_i+t)}\phi^{(0)}_{i-1}.\label{eqn:time_jump}
\end{eqnarray}
We insert unity as $1=\sum_n |\phi_i^{(n)}\rangle\langle\phi_i^{(n)}|$ into the above equation to find
\begin{eqnarray}
\psi=\sum_{n}{e^{-E_i^{(n)}(t_i+t)}\langle\phi^{(n)}_{i}|\phi^{(0)}_{i-1}\rangle\phi^{(n)}_{i}}.
\end{eqnarray}
Note that for $n$ odd, the overlap vanishes since $\phi^{(0)}_{i-1}$ is an even function. To keep the probabilistic interpretation of $e^{-\beta U/2}\psi$, the normalization is chosen as 
\begin{equation}
\int \exp\left(-\frac{\beta_i x^2}{4}\right)\psi dx=1.
\end{equation} Therefore, the normalized ground state solution of the Hamiltonian $H_{i-1}$ reads
\begin{equation}\label{eqn:phi_0}
\phi^{(0)}_{i-1}=\left(\frac{\beta_{i-1}}{2\pi}\right)^{1/2}\exp\left(-\frac{\beta_{i-1} x^2}{4}\right)
\end{equation}
 and the new wave function up to the slowest decaying term is 
\begin{eqnarray}
\psi= \frac{\sqrt{\beta_i\beta_{i-1}}}{\displaystyle{\frac{\beta_i+\beta_{i-1}}{2}}}\phi^{(0)}_{i}+\langle\phi^{(2)}_{i}|\phi^{(0)}_{i-1}\rangle e^{-E_i^{(2)}(t_i+t)}\phi^{(2)}_{i}\label{psievol}.
\end{eqnarray}
For our approach to be sensible we have to wait again long enough for the decay of the excited state $\phi_i^{(2)}$. However, the decay constant $\tau_i=1/E_i^{(2)}$, as we learn from Eq. (\ref{eqn:E_i}), is exponentially increasing with $\beta$, and therefore, we will have to wait longer and longer as time proceeds for the system to decay.

Now, we set $H_i\rightarrow H_i+\lambda(1-P_i^{(0)})$, with $P_i^{(0)}$, the projector to the ground state. Then Eq. (\ref{psievol}) becomes
\begin{eqnarray}
\psi= \frac{\sqrt{\beta_i\beta_{i-1}}}{\displaystyle{\frac{\beta_i+\beta_{i-1}}{2}}}\phi^{(0)}_{i}+\langle\phi^{(2)}_{i}|\phi^{(0)}_{i-1}\rangle e^{-(E_i^{(2)}+\lambda)(t_i+t)}\phi^{(2)}_{i}\label{psievol.lambda}.
\end{eqnarray}
For any monotonically growing $\beta$, $E_i^{(2)}$ vanishes exponentially, so that the slowest decaying terms of Eq. (\ref{psievol.lambda}) decay approximately as $e^{-\lambda (t_i+t)}$. Therefore, the decay constant is bounded from below by $\tau_i=1/\lambda$ and is thus independent of the index $i$.

Let us investigate what the above considerations mean for the average energy at a time $t>t_i$, shortly before the next jump of $\beta(t)$ at $t_{i+1}$,
\begin{eqnarray}
\langle E\rangle(t)&\approx&\int dx\frac{x^2}{2}\exp\left(-\frac{\beta_i x^2}{4}\right)\frac{\sqrt{\beta_i\beta_{i-1}}}{\displaystyle{\frac{\beta_i+\beta_{i-1}}{2}}}\phi^{(0)}_{i}\nonumber\\
&=&\frac{\sqrt{\beta_i\beta_{i-1}}}{\displaystyle{\frac{\beta_i+\beta_{i-1}}{2}}}\int dx \frac{x^2}{2}\exp\left(-\frac{\beta_i x^2}{4}\right)\left(\frac{\beta_i}{2\pi}\right)^{1/2}\exp\left(-\frac{\beta x^2}{4}\right)\nonumber\\
&=& \left(\frac{\beta_i}{2\pi}\right)^{1/2}\frac{\sqrt{\beta_i\beta_{i-1}}}{\displaystyle{\frac{\beta_i+\beta_{i-1}}{2}}}
\frac{\sqrt{2\pi}}{\beta_i^{3/2}}\nonumber\\
&=&\frac{\sqrt{\beta_i\beta_{i-1}}}{\displaystyle{\frac{\beta_i+\beta_{i-1}}{2}}}\frac{1}{\beta_i},
 \end{eqnarray}
 where we have used Eq. (\ref{eqn:phi_0}).
For the schedule $\beta_i=\beta_0+\delta i$ the prefactor becomes 
\begin{eqnarray} \frac{\sqrt{\beta_i\beta_{i-1}}}{\displaystyle{\frac{\beta_i+\beta_{i-1}}{2}}}=\frac{2\sqrt{(\beta_0+ki)(\beta_0+k(i-1))}}{2\beta_0+2ki-k}\approx\frac{2k\sqrt{i(i-1)}}{k(2i-1)}\rightarrow 1\quad\text{for $i\gg 1$}.
\end{eqnarray}
Thus the average energy scales as $\langle E\rangle\sim 1/{\delta i}$, in contrast to $\langle E\rangle\sim B/{\ln (i)}$, as is the case for the logarithmic schedule.
\subsubsection{Traveling Salesman Problem} 
\begin{figure}[ht]
\includegraphics[width=\textwidth]{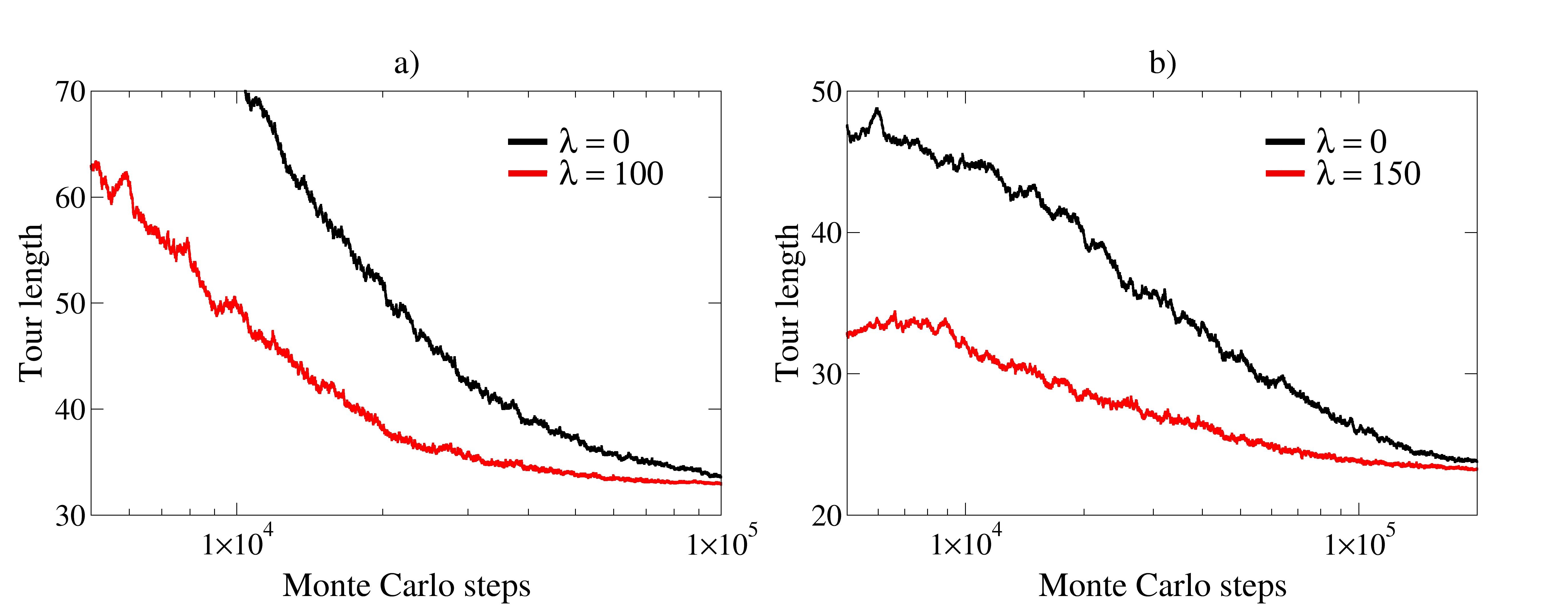} 
\caption{(Color online) Results of simulated annealing for the traveling salesman problem: a) `gr666' and b) `u1060' data sets with 666 and 1060 cities, respectively.} \label{SATSP}
\end{figure}

We now turn our attention to the traveling salesman problem. We use the `gr666' and `u1060' data sets from the TSP-database \cite{reinelt} with 666 and 1060 cities, respectively. The inverse temperature is chosen to be stepwise growing with Monte Carlo steps. Starting from a base value $\beta_0=40$, the inverse temperature is increased by $\delta =3.5$ after 3000 Monte Carlo steps, similarly to the plot in Fig. \ref{fig:schedule}.
For the simulation we used the simplified method described in Sec. \ref{sec:implementation}.
At each Monte Carlo step, we chose $N$ nearest neighbors and $N$ non-local states \citep{lin} randomly, where $N$ is the number of cities.
The results of the simulations are shown in Fig. \ref{SATSP} a) for `gr666' and b) `u1060', where the tour length is plotted versus the logarithm of the Monte Carlo steps. The effect and advantage of using our method are clearly visible. Sampling for the same amount of iterations, we find much better solutions by using the transition rates modified by $\lambda$, as defined in Eq. (\ref{eqn:out}), than when using $\lambda=0$. Stated otherwise, we can achieve an acceleration by an order of magnitude to reach a given tour length.

\subsubsection{Three-dimensional Gaussian Ising Spin Glass}\label{sec:3dimsa}
\begin{figure}[hb]
\includegraphics[width=\textwidth]{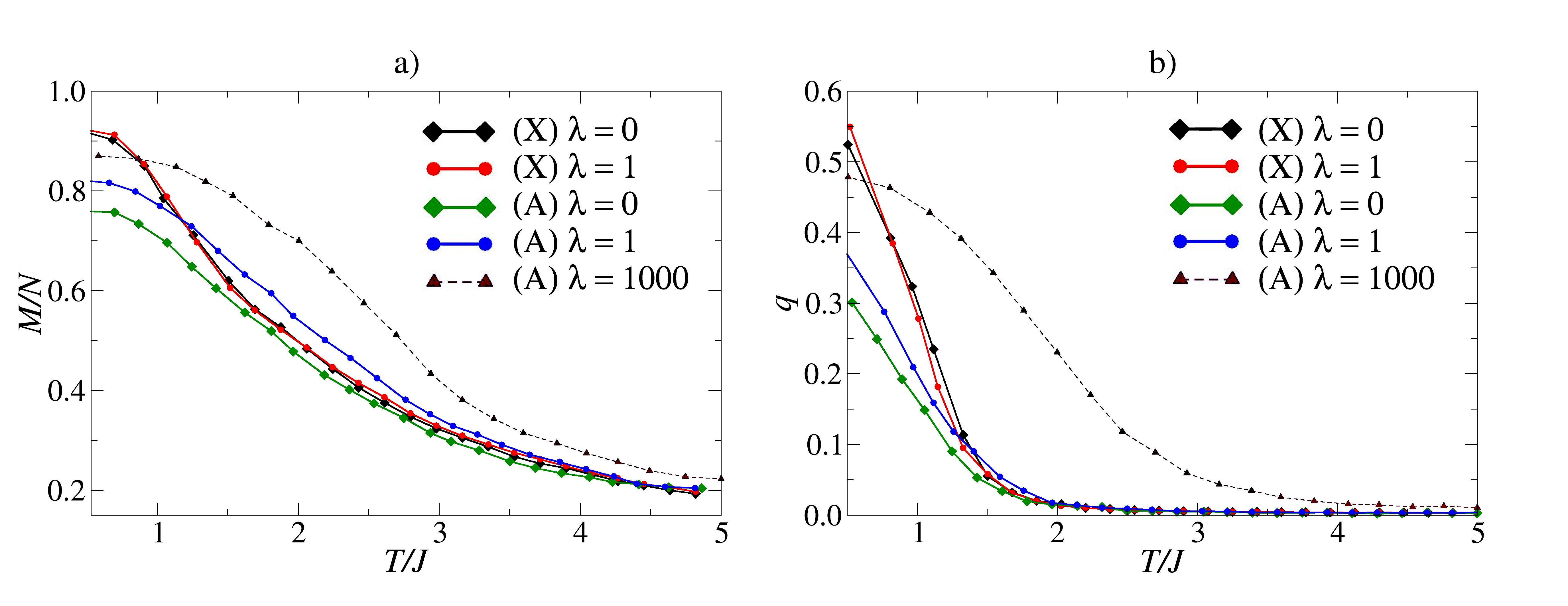} 
\caption{(Color online) Comparison of the results obtained by simulated annealing (A) and exchange Monte Carlo (X) of the three-dimensional Ising spin glass. a) Magnetization per spin b) Edwards-Anderson order parameter. Note that the finite-temperature values of the present method (curves marked (A), $\lambda>0$) do not correspond to equilibrium properties.} \label{XMC}
\end{figure}

Next we treat the three-dimensional Gaussian Ising spin glass (see Sec. \ref{sec:models}). It is clear that a na\"ive simulated annealing, with local jumps only, fails to find a good low temperature solution of this model in a reasonable time. There are many energetically close states which are separated by huge barriers, and thus the time needed to escape from a local minimum increases very quickly as the temperature decreases. Nevertheless, we would like to see how our method affects the annealing procedure. The temperature is lowered in 480 steps from $T_{\rm{max}} = 5/J$ to $T_{\rm{min}}=0.5/J$. The coupling constants of the Hamiltonian are drawn from a Gaussian distribution, with zero mean and variance $J$. The external field strength is chosen as $h/J=0.1$. In Fig. \ref{XMC} we compare the a) magnetization and b) Edwards-Anderson order parameter, $ q=\left\langle (\frac{1}{N}\sum_i S^{(1)}_iS^{(2)}_i )^2\right\rangle $,
of simulated annealing (A) with results from exchange Monte Carlo calculations (X). 
Notice that the finite-temperature values of the present method, marked (A), do not represent the equilibrium properties for the reason discussed in Sec. \ref{sec:implementation}. The exchange Monte Carlo is usually expected to give good results although we have not checked equilibration conditions since our goal is to compare the performance of the methods under the same conditions on computational cost.

The first observation from the data is that the introduction of the $\lambda$ term significantly improves the performance of simulated annealing at the lowest temperature. The values of physical quantities, $M/N$ and $q$, have come close to those of the exchange Monte Carlo, the latter being a benchmark.
Another notable fact is that the $\lambda$ term, at least for a small value, induces no perceptible change in the exchange Monte Carlo. Lastly, the very large value of $\lambda=1000$ yields close results to those of the exchange Monte Carlo at the lowest temperature. The results for $\lambda=1000$ are, nevertheless, still slightly away from the exchange Monte Carlo values. These facts clarify the usefulness as well as limits of the present method for this problem of Ising spin glass.

\section{Conclusion}\label{sec:conclusio}
Based on the idea of Ref. \citep{orland}, we introduced a method to accelerate stochastic sampling of discrete, classical problems. Our method suggests a way to overcome the limits of standard simulated annealing. We tested our algorithm on the traveling salesman problem, where, in the framework used in this paper, we find the shortest tour an order of magnitude faster by using our method than in the conventional case. Simulated annealing of the three-dimensional Gaussian Ising spin glass is also accelerated. In this latter case, the performance of our method is relatively close to that of the exchange Monte Carlo.

 Throughout our investigation we used a simplified version of the full kinetic Monte Carlo algorithm to reduce the computational cost of the sampling at each Monte Carlo step. This simplified method outperforms the full kinetic Monte Carlo when we are faced with a plethora of accessible states, but when the choices are limited, adding $\lambda$ tends to have undesirable effects if it is not chosen accordingly. However, in its present form the algorithm is useful only in the search for very low temperature solutions. 

In conclusion, the present method would be a useful alternative of simple simulated annealing for optimization problems. The relatively straightforward implementation using kinetic Monte Carlo and non-local moves would make it a method of choice for some purposes, especially where computational cost is a factor.

\acknowledgements
We thank Koji Hukushima for providing us with data for comparison of exchange Monte Carlo calculations. Z.B. thanks Yoshiki Matsuda for useful comments. This work was supported by CREST, JST.


\begin{thebibliography}{40}
\bibitem{hartmann-rieger} A. K. Hartmann and H. Rieger, \textit{Optimization Algorithms in Physics}, (Wiley, Berlin, 2002).
\bibitem{metropolis} N. Metropolis, A. W. Rosenbluth, M. N. Rosenbluth, A. H. Teller and E. Teller, \textit{J. Chem. Phys.} \textbf{21}, 1087 (1953).
\bibitem{voter} A. F. Voter, in \textit{Radiation Effects in Solids, K.E. Sickafus and E.A. Kotomin (eds.)}, (Springer, Dordrecht, 2005).
\bibitem{KGV} S. Kirkpatrick, C. D. Gelatt and M. P. Vecchi, \textit{Science} \textbf{222}, 220 (1983).	V. Cerny, \textit{J. Opt. Theor. App.} \textbf{45}, 41 (1985).
\bibitem{XMC} K. Hukushima and K. Nemoto, \textit{J. Phys. Soc. Jpn.} \textbf{65}, (1996).
\bibitem{popa} K. Hukushima and Y. Iba, in \textit{The Monte Carlo Method in the Physical Sciences, J.E. Gubernatis (ed.)}, (The American Institute of Physics, 2003).
\bibitem{pfolding} A. R. Fersht and V. Daggett, \textit{Cell} \textbf{108}, 573 (2002).
\bibitem{orland} H. Orland, \textit{J. Phys. Soc. Jpn.} \textbf{78}, 103002 (2009).
\bibitem{reinelt} G. Reinelt, \textit{ORSA J. Comp.} \textbf{3}, 376 (1991). \url{http://comopt.ifi.uni-heidelberg.de/software/TSPLIB95 }
\bibitem{shinokaba} S. Shinomoto and Y. Kabashima, \textit{J. Phys. A: Math. Gen.} \textbf{24}, L141 (1991).
\bibitem{geman} S. Geman and D. Geman, \textit{IEEE Trans. Patt. Anal. Mach. Intel.} \textbf{6}, 723 (1984).
%\bibitem{reichl} L. E. Reichl, \textit{A Modern Course in Statistical Physics 2nd 
%Ed.}, (John Wiley and Sons Inc., New York, 1998).
%\bibitem{jarzyinski} C. Jarzyinski, \textit{Phys. Rev. Lett.} \textbf{78} 2690 (1997).
\bibitem{johnson} D. S. Johnson and L. A. McGeoch, in \textit{Local Search in Combinatorical Optimization, E.H.L. Aarts and J.K. Lenstra (eds.)}, (John Wiley and Sons Inc., London, 1997).
\bibitem{lin} S. Lin and B. W. Kernighan, \textit{Oper. Res.} \textbf{21}, 498 (1973).
\bibitem{wolff}  R. H. Swendsen and J. S. Wang, \textit{Phys. Rev. Lett.} \textbf{58}, 86 (1987). U. Wolff, \textit{Phys. Rev. Lett.} \textbf{62}, 361 (1989).
\bibitem{talapov} A. L. Talapov and H. W. J. Bl\"ote, \textit{J. Phys A: Math. Gen.} \textbf{29}, 5727 (1996).
\bibitem{schwabl} G. Schwabl, \textit{Quantum Mechanics}, (Springer, Heidelberg, 1995).
%\bibitem{hartmann} A. K. Hartmann, \textit{Physica} A \textbf{224} 480 (1996).




\end{thebibliography}
\end{document}